\documentclass[a4paper,11pt]{article}

\pdfoutput=1
\usepackage[paper=a4paper,margin=1.25in]{geometry}

\usepackage[hyperfootnotes=false, linktocpage=true, colorlinks, citecolor=blue, linkcolor=blue, urlcolor=blue, filecolor=blue]{hyperref}
\usepackage{graphicx}
\usepackage{amsmath}
\usepackage{amssymb}
\usepackage{cite}
\usepackage{color}
\usepackage{enumitem}
\usepackage{xspace}
\usepackage{array}
\usepackage{subfig}

\usepackage{epsfig}
\usepackage{bbm}

\usepackage{multirow} % merger multi cells in a table

\numberwithin{equation}{section}
\newlength{\xtrawidth}
\newlength{\xtraheight}
\def\C{\mathbb{C}}

\def\C{\mathcal{C}}

\newcommand{\beq}{\begin{equation}}
\newcommand{\eeq}{\end{equation}}

\newcommand{\shortline}{\newline\vskip -7mm{\hbox to 2cm{\hrulefill}}\vskip 3mm}

\newcommand{\OP}{{\cal O}_{\mathbb{P}^1}}

\newcommand{\qed}{\nobreak \ifvmode \relax \else
      \ifdim\lastskip<1.5em \hskip-\lastskip
      \hskip1.5em plus0em minus0.5em \fi \nobreak
      \vrule height0.75em width0.5em depth0.25em\fi}

\newcommand{\be}{\begin{equation}} % start the equation
\newcommand{\ee}{\end{equation}} % end the equation
\newcommand{\bi}{\begin{itemize}} % start the item env
\newcommand{\ei}{\end{itemize}} % end the item env 

\renewcommand*{\thefootnote}{\fnsymbol{footnote}} % to present authors information

\begin{document}
\begin{centering}
\vspace*{1.2cm}
{\LARGE \bf Instantons and Hilbert Functions}

\vspace{1cm}

{\bf Evgeny I. Buchbinder}${}^{1,}$\footnote{evgeny.buchbinder@uwa.edu.au},
{\bf Andre Lukas}${}^{2,}$\footnote{lukas@physics.ox.ac.uk},
{\bf Burt A. Ovrut}${}^{3,}$\footnote{ovrut@elcapitan.hep.upenn.edu},
{\bf Fabian Ruehle}${}^{4,2,}$\footnote{fabian.ruehle@cern.ch}

{\small
\vspace*{.5cm}
${}^{1}$Department of Physics, The University of Western Australia,\\
35 Stirling Highway, Crawley WA 6009, Australia\\ [3mm]
${}^{2}$Rudolf Peierls Centre for Theoretical Physics, University of Oxford\\
  Parks Road, Oxford OX1 3PU, UK\\[3mm]
${}^{3}$Department of Physics and Astronomy, University of Pennsylvania\\Philadelphia PA 19104-6396, USA\\[3mm]
${}^{4}$CERN, Theoretical Physics Department\\
1 Esplanade des Particules, Geneva 23, CH-1211, Switzerland
}

\begin{abstract}\noindent
We study superpotentials from worldsheet instantons in heterotic Calabi-Yau compactifications for vector bundles constructed from line bundle sums, monads and extensions. Within a certain class of manifolds and for certain second homology classes, we derive simple necessary conditions for a non-vanishing instanton superpotential. These show that non-vanishing instanton superpotentials are rare and require a specific pattern for the bundle construction. For the class of monad and extension bundles with this pattern, we derive a sufficient criterion for non-vanishing instanton superpotentials based on an affine Hilbert function. This criterion shows that a non-zero instanton superpotential is common within this class. The criterion can be checked using commutative algebra methods only and depends on the topological data defining the Calabi-Yau $X$ and the vector bundle $V$.
\end{abstract}
\end{centering}

\newpage

\tableofcontents
\setcounter{footnote}{0}
\renewcommand*{\thefootnote}{\arabic{footnote}}  % to use numerical footnotes again

%%%%%%%%%%%%%%%%%%%%%%%%%%%%%%%%%%%%%%%%%%%%%%%%%%%%%%%%%%%%%%%%%%%%%%%%%%%%%

\section{Introduction}
Non-perturbative superpotentials generated from instanton effects play an important role in string 
theory~\cite{Dine:1986zy,Dine:1987bq,Becker:1995kb,Witten:1996bn,Donagi:1996yf,Witten:1999eg,Harvey:1999as,Lima:2001jc,Lima:2001nh,Buchbinder:2002ic,Buchbinder:2002pr} 
and they form a crucial ingredient for a stability analysis of string vacua and for practically all scenarios of moduli stabilization. It is well-known that contributions to the instanton superpotential are proportional to $\exp(-{\rm Vol}(C))$, where $C$ is the (calibrated) cycle wrapped by the string or the brane. However, more detailed calculations including the pre-factor of this exponential are often difficult to carry out and explicit results are few and far between.  In particular, it is not easy to determine whether the instanton superpotential is zero or non-zero.\\[2mm]
In this paper, we are concerned with superpotentials from string worldsheet instantons in heterotic compactifications  on Calabi-Yau three-folds $X$ with vector bundles $V\rightarrow X$. For such compactifications, the instanton superpotential can receive a contribution $W_\C\sim \exp(-{\rm Vol}(\C))$ from each second homology class $\C$, where all isolated, genus zero holomorphic curves $C_i$, $i=1,\ldots ,n_\C$, in the class $\C$ contribute to the pre-factor in $W_\C$.\\[2mm]
Beasley and Witten~\cite{Beasley:2003fx} have studied linear and half-linear sigma models and have shown that the contributions from the curves $C_i$ sum up to zero, and, hence, that $W_\C$ vanishes, under fairly general assumptions (see also Refs.~\cite{Distler:1986wm,Distler:1987ee,Silverstein:1995re,Basu:2003bq}).
On the other hand, a number of papers~\cite{Buchbinder:2016rmw,Buchbinder:2017azb,Buchbinder:2018hns,Buchbinder:2019hyb} have produced examples with a non-vanishing $W_\C$, thus apparently evading the vanishing theorems of Ref.~\cite{Beasley:2003fx}. There are two obvious resolutions: Either there is a problem with the geometric methods used to calculate the instanton contributions or the examples considered violate one of the assumptions underlying the vanishing theorems of Beasley and Witten. 
The results of Ref.~\cite{Buchbinder:2019hyb}, points to the latter being the correct explanation.\\[2mm]
More specifically, one of the assumptions underlying the vanishing theorems is compactness of the instanton moduli space. Unfortunately, this assumption is not easily checked in general. A nice  straightforward method, due to Bertolini and Plesser~\cite{Bertolini:2014dna}, is only available if a GLSM formulation of the model can be found. This limits the models for which the assumptions can be checked with this method and requires, among other things, that the bundle $V$ is given as a monad bundle. However, 
in Ref.~\cite{Buchbinder:2019hyb} the authors have identified a number of models for which the geometric calculation can be carried out and a GLSM formulation can be found. In all those cases, the result of the geometric calculation turn out to be consistent with the vanishing theorems, as formulated by Bertolini and Plesser.\\[2mm]
In the present paper, we would like to invert the logic and assume, based on the evidence in Ref.~\cite{Buchbinder:2019hyb}, that the standard geometric methods to calculate instanton superpotential are indeed correct and consistent with the vanishing theorems. By applying these methods we would like to address two main questions.
\begin{itemize}
 \item Can we find simple conditions for the vanishing/non-vanishing of the instanton superpotential based solely on the geometric data $(X,V)$? These can be thought of as  geometric analogues of the Bertolini-Plesser GLSM conditions, but applicable to a wider class of models for which no GLSM description is known.
 \item How ``common" is it for the instanton superpotential to be vanishing or non-vanishing? 
\end{itemize} 
As we will see, the first question can be partially answered in terms of simple cohomology conditions and a certain affine Hilbert function which we introduce. Analyzing these, we find that a non-vanishing instanton superpotential only arises within a specific sub-class of bundles $V$, but that it is common to be non-vanishing  within this sub-class.\\[2mm]
The plan of the paper is as follows. In the next section, we review the standard geometric method to calculate string instanton superpotentials. As we will see, this method requires explicit knowledge of the isolated, genus zero curves $C_i$, which can be difficult to determine explicitly. In Section~\ref{sec:CY}, we introduce a class of (complete intersection) Calabi-Yau manifolds where these curves can be found, at least for certain homology classes $\C$. Basic features of common vector bundle constructions, including line bundle sums, monad bundles and extension bundles, relevant for our discussion of instantons, are summarized in Section~\ref{sec:bundle}. The requisite mathematical background on coordinate rings and Hilbert functions is reviewed in Section~\ref{sec:hilbert}. In Section~\ref{sec:crit}, we formulate the Hilbert function criterion for non-vanishing instanton superpotentials and apply it to a number of examples. We conclude in Section~\ref{sec:conclusion}.

\section{Geometric calculation of instanton superpotentials}\label{sec:general}
In this section, we first review a method for calculating instanton superpotentials based on techniques from algebraic geometry (see, for example, Refs.~\cite{Lima:2001jc,Buchbinder:2002ic,Buchbinder:2002pr} for more details). 

We are working in the context of $E_8\times E_8$ heterotic string compactifications on Calabi-Yau three-folds to four-dimensional theories with $\mathcal{N}=1$ supersymmetry. Our main object of interest is the superpotential of the four-dimensional theory generated by string instanton effects. 

The basic data which defines the compactification consist of a Calabi-Yau three-fold $X$ and a holomorphic, poly-stable vector bundle $V\rightarrow X$ with $c_1(V)=0$ and a structure group which can be embedded into $E_8$. In general, there is also another bundle whose structure group embeds into the second $E_8$ factor and/or five branes wrapping holomorphic curves in $X$. Details of these further ingredients are not really relevant for our discussion but we would like to ensure that there exist choices of a second bundle or five-branes such that the compactification is anomaly-free and respects supersymmetry. This is guaranteed if we demand that the curve dual to $c_2(TX)-c_2(V)$ is an element of the Mori cone of $X$ for a poly-stable $V$. In this case, an anomaly-free, supersymmetric completion can, for example, be achieved by wrapping five-branes on a holomorphic curve with class $c_2(TX)-c_2(V)$.\\[2mm]
The instanton superpotential $W$ in the resulting four-dimensional theory can be written as a sum $W=\sum_\C W_\C$ over contributions $W_\C$ which are associated to classes $\C\in H_2(X,\mathbb{Z})$ in the second homology of $X$. We will usually focus on one of these homology classes $\C$ and will attempt to compute $W_\C$. The superpotential term $W_\C$ receives contributions from the isolated, genus zero holomorphic curves with class $\C$. We denote these curves by $C_i$, where $i=1,\ldots ,n_\C$ and $n_\C$ is the genus zero Gromov-Witten invariant. Schematically, the superpotential term $W_\C$ can be written as
\begin{equation}
 W_\C=\left[\sum_{i=1}^{n_\C}{\rm Pfaff}_{C_i}\right]\exp\left(-\int_\C (J+iB)\right)\label{instsp}
\end{equation} 
where $J$ is a K\"ahler form on $X$, $B$ is the NS two-form and ${\rm Pfaff}_{C_i}$ is the Pfaffian. Its precise form in terms of differential operators on the curve $C_i$ can be found, for example, in Ref.~\cite{Witten:1999eg}. The instanton superpotential associated to the class $\C$ is, of course, proportional to the exponent $\exp(-{\rm Vol}(\C))$. The (one-loop) pre-factor in Eq.~\eqref{instsp} corresponds to the various contributing isolated, genus zero curves $C_i$ with class $\C$ which are wrapped by instantonic strings. \\[2mm]
From a theoretical perspective as well as in the context of physical applications, such as for example in applications to moduli stabilization, it is crucial to know whether the pre-factor $\sum_i{\rm Pfaff}_{C_i}$ in Eq.~\eqref{instsp} is zero or non-zero. This is the main question we will address in the present paper. 

How can the Pfaffians ${\rm Pfaff}_{C_i}$ be computed in practice? The key statement~\cite{Buchbinder:2002ic}  underlying the algebraic computation is formulated in terms of the bundle
\begin{equation}
 V_i:=V|_{C_i}\otimes{\cal O}_{\mathbb{P}^1}(-1)
\end{equation} 
and asserts the following equivalence.
\begin{equation}
 H^0(V_i)\neq 0\qquad\Longleftrightarrow\qquad {\rm Pfaff}_{C_i}=0\; . \label{H0meth}
\end{equation} 
Broadly speaking, the idea is to work out the cohomology on the left-hand side, rather than computing the Pfaffian directly. More specifically, we note that the value of this cohomology does depend on the choice of moduli, that is, on the complex structure moduli of $X$ and on the bundle moduli of $V$. Here we will generally assume that the complex structure moduli of $X$ have been fixed to suitably generic values and focus on the dependence on the bundle moduli of $V$, which we denote as $b=(b_\alpha)$. Of course it is possible that the cohomology in \eqref{H0meth} is non-zero for all values of $b$. In this case, the Pfaffian, as a function of $b$, vanishes identically.\\[2mm]
A more interesting situation arises when the cohomology in \eqref{H0meth} vanishes for generic values of $b$ but has a ``jumping locus" in bundle moduli space where it acquires a non-zero value. As we will see, such a jumping locus is described by an equation of the form $f_i(b)=0$, where $f_i$ is a holomorphic function. Since this function $f_i$ and the Pfaffian ${\rm Pfaff}_{C_i}$ have an identical zero locus they must be proportional. Hence, we can write
\begin{equation}
 W_\C=\left[\sum_{i=1}^{n_\C}\lambda_i f_i(b)\right]\exp\left(-\int_\C (J+iB)\right)\; ,\label{instsp1}
\end{equation} 
where $\lambda_i\in\mathbb{C}$ are constants.\\[2mm]
Unfortunately, we do not currently know how to compute the constants $\lambda_i$ in Eq.~\eqref{instsp1}, at least not with algebraic methods. In fact, these constants are tied up with a rather subtle interpretation~\cite{Witten:1999eg} of the NS two-form field $B$. Unfortunately, our ignorance in this respect somewhat obstructs our ability to answer the  question about the vanishing of $W_\C$. Luckily, not all is lost if the $f_i$ are indeed non-trivial functions of the moduli $b$, as is frequently the case. Then we have
\begin{equation}
 (f_i)_{i=1,\ldots ,n_\C}\; \mbox{linearly independent functions}\qquad\Longrightarrow\qquad W_\C\neq 0\; . \label{fcrit}
\end{equation} 
This is the basic criterion which will underlie much of our discussion. It allows for a definite conclusion if the functions $f_i$ are linearly independent - in this case $W_\C$ is a non-zero function. If the $f_i$ are linearly dependent the answer depends on the unknown constants $\lambda_i$. If their values are such that they realise the linear dependence relation $\sum_i\lambda_if_i=0$ then $W_\C$ vanishes, otherwise $W_\C$ is still non-zero.\\[2mm]
Any computation along the above lines requires, in a first instance, explicit knowledge of the isolated, genus-zero curves\footnote{We focus on the dominant instanton contributions, which arise from curves with single wrapping.} $C_i$ in a given class $\C$. Finding these curves can be quite non-trivial, so any concrete progress depends on a setting where these curves can be found. We will now review how this can be done for a certain class of Calabi-Yau manifolds.

\section{The Calabi-Yau manifolds} \label{sec:CY}
\subsection{General set-up} \label{sec:setup}
We consider an ambient space of the form ${\cal A}=\mathbb{P}^1\times{\cal B}$, where ${\cal B}=\mathbb{P}^{n_1}\times\cdots\times\mathbb{P}^{n_m}$, with homogeneous coordinates $x=(x_0,x_1)$ for the $\mathbb{P}^1$ factor and $y=(y_{\alpha,0},\ldots ,y_{\alpha,n_\alpha})_{\alpha=1,\ldots ,m}$ for the other factors. In this ambient space, we define complete intersection Calabi-Yau manifolds (CICYs) $X$ which are specified by a configuration matrix
\begin{equation}
\mbox{type I}:\; X\in\left[\begin{array}{l|lllll}\mathbb{P}^1&1&1&0&\cdots&0\\
                                          {\cal B}&q_1&q_2&q_3&\cdots&q_K
              \end{array}\right]\; ,\qquad\mbox{type II}:\;
 X\in\left[\begin{array}{l|llll}\mathbb{P}^1&2&0&\cdots&0\\
                                          {\cal B}&q_1&q_2&\cdots& q_K
              \end{array}\right] \; .           
\end{equation}                                          
Every column of the configuration matrix indicates the multi-degree of a homogeneous polynomial $P_a=P_a(x,y)$ and the CICY manifold $X$ is the common zero locus of these polynomials. The Calabi-Yau condition, $c_1(TX)=0$, is equivalent to the degrees in each row of the configuration matrix summing up to the dimension of the projective space plus one. For $\mathbb{P}^1$ this leaves only two possible patterns for the degree and this is how the above two types arise. \\[2mm]
The point about these CICY manifolds, as shown in Ref.~\cite{Buchbinder:2017azb}, is that the isolated, genus zero curves in the class $\C$ which corresponds to the first $\mathbb{P}^1$ factor can be determined rather straightforwardly. We briefly review how this works, starting with type I. In this case, the defining polynomials can be written as
\begin{equation}
 \begin{array}{lll}
 P_1(x,y)&=&x_0\, Q_1(y)+x_1\, Q_2(y)\\
 P_2(x,y)&=&x_0\, Q_3(y)+x_1\, Q_4(y)\\
 P_a(x,y)&=&Q_{a+2}(y)\;\qquad\qquad\qquad\mbox{for}\; a>2\; ,
\end{array} 
\end{equation} 
where $Q_1$ and $Q_2$ have multi-degree $q_1$, $Q_3$ and $Q_4$ have multi-degree $q_2$ and $Q_{a+2}$ for $a>2$ has multi-degree $q_a$. For type II, the analogous decompositions are
\begin{equation}
 \begin{array}{lll}
  P_1(x,y)&=&x_0^2\,Q_1(y)+x_0x_1\,Q_2(y)+x_1^2\,Q_3(y)\\
  P_a(x,y)&=&Q_{a+2}(y)\qquad\qquad\qquad\qquad\qquad\qquad\;\mbox{for}\; a>2
  \end{array}
\end{equation}
where $Q_1$, $Q_2$ and $Q_3$ have multi-degree $q_1$ and and $Q_{a+2}$ for $a>2$ has multi-degree $q_a$. For either type, the defining equations $P_1(x,y)=\cdots =P_K(x,y)=0$ of the CICY manifold are solved for all $x\in\mathbb{P}^1$ if
\begin{equation}
 Q_1(y)=Q_2(y)=\cdots =Q_{K+2}(y)=0\; . \label{Qeqs}
 \end{equation}
These last equations define a zero-dimensional complete intersection in the space ${\cal B}$ which corresponds to a finite number of points $Y_i$. This finite point set can also be represented by the configuration matrices
\begin{equation}
\begin{array}{rl}
 \mbox{type I}:& \{Y_i\}\in\left[\begin{array}{l|lllllll}{\cal B}&q_1&q_1&q_2&q_2&q_3&\cdots&q_K\end{array}\right]\\[3mm]
  \mbox{type II}:& \{Y_i\}\in\left[\begin{array}{l|llllll}{\cal B}&q_1&q_1&q_1&q_2&\cdots&q_K\end{array}\right]\; .
\end{array}  
\end{equation} 
In this way, we have identified a number of isolated, genus-zero curves $\mathbb{P}^1\times Y_i\subset X$, where $i=1,\ldots ,n_\C$ in the class $\C$ associated to the first $\mathbb{P}^1$ factor. By computing the Gromov-Witten invariant for this class~\cite{Buchbinder:2017azb}, it can be shown that this is indeed the complete set of such curves.\\[2mm]
For the calculation of instanton superpotentials along the lines described in Section~\ref{sec:general}, we need to find the isolated, genus-zero curves explicitly. The above set-up presents us with a straightforward way to do this by solving the equations~\eqref{Qeqs} for the loci $Y_i$ of these curves in the ``transverse" space ${\cal B}$. Note that, while this is conceptually simple, it can still be very hard to carry out in practice. Finding the exact solutions to Eqs.~\eqref{Qeqs} is impossible for anything but the simplest cases and even numerical solutions can be difficult to come by. The alternative algebraic approach we will be formulating is circumventing this problem - it requires no explicit knowledge of the points $Y_i$.\\[2mm]
Finally, we introduce an algebraic descriptions of the above set-up. The point set $\{Y_i\}$ is a zero-dimensional algebraic variety but there are two, subtly different ways to think about this. For one, we can think of $\{Y_i\}$ as a projective sub-variety of ${\cal B}$ and associate to it the  projective ideal
\begin{equation}
 I=\langle Q_1,\ldots ,Q_{K+2}\rangle\; .
\end{equation} 
Alternatively, we can also think about the point set $\{Y_i\}$ as an affine variety. To this end, we focus on the patch $U_0$ of ${\cal B}$ where all $y_{\alpha,0}\neq 0$ and we assume that the defining polynomials $Q_a$ are sufficiently generic such that all points $Y_i$ are contained in $U_0$. Then, we can think of the point set $\{Y_i\}$ as an affine sub-variety of $U_0$ and associate to it an ideal $J$ which is obtained from $I$ by adding the ``localising" generators $y_{\alpha,0}-1$. Hence, $J$  is explicitly given by
\begin{equation}
 J=\langle Q_1,\ldots ,Q_{K+2},y_{1,0}-1,\ldots ,y_{m,0}-1\rangle\; .
\end{equation}
Associated to the ideals $I$ and $J$ are projective and affine coordinate rings, respectively, and we have the following maps between those rings:
\begin{equation}
 \label{SAmap}
 \mathbb{C}[y]\stackrel{r}{\longrightarrow} S\stackrel{\ell}{\longrightarrow} A\qquad\text{with}\qquad S:=\frac{\mathbb{C}[y]}{I}\,,~~ A:=\frac{\mathbb{C}[y]}{J}\,.
\end{equation} 
Here, $r$ maps a polynomial in $\mathbb{C}[y]$ to its associated class in $S$ and $\ell$ is a localisation map, effectively carried out by setting all $y_{\alpha,0}=1$. Note that the affine ring $A$ is, in fact, finite-dimensional with dimension equal to $n_\C$, the number of points $Y_i$. \\[2mm]
As we will see, these algebraic descriptions of the curve loci $\{Y_i\}$ in terms of coordinate rings are key to our subsequent discussion of instantons. In particular, the rings $S$ and $A$ do not explicitly depend on the points $Y_i$ but merely on the polynomials $Q_a$. This feature means that out algebraic approach will not rely on the explicit knowledge of these points.

\subsection{A few simple examples} \label{sec:examples}
It is useful to introduce a few simple examples which can be used to illustrate our method as we go along. We emphasise that the following examples are specifically chosen for their simplicity, particularly a small number, $n_\C$, of curves, so that an explicit ``on paper" treatment is possible. Our method will of course not be restricted to such simple cases and some more complicated examples will be described later.\\[2mm]
{\bf Example 1:} {\it A type I example with two projective factors}\\
Consider the CICY manifold $X$ (number $7867$ in the standard list~\cite{Candelas:1987kf,Green:1986ck}) with configuration matrix
\begin{equation} 
 X\in\left[\begin{array}{l|llll}\mathbb{P}^1&0&0&1&1\\\mathbb{P}^6&3&2&1&1\end{array}\right]^{2,68}_{-132}\qquad
 \begin{array}{l}x_0,x_1\\y_0,\ldots ,y_6\end{array} \label{confex1}
\end{equation} 
where the Hodge numbers $h^{1,1}(X),h^{2,1}(X)$ are attached as a superscript and the Euler number as a subscript. The single-wrapping Gromov-Witten invariant associated to the class of the $\mathbb{P}^1$ factor is $n_\C=6$ and the configuration matrix specifying the six loci $Y_i$ of these curves in the transverse space $\mathbb{P}^6$ is
\begin{equation}
 \{Y_1,\ldots ,Y_6\}\in \left[\begin{array}{l|llllll}\mathbb{P}^6&3&2&1&1&1&1\end{array}\right]\cong\left[\begin{array}{l|ll}\mathbb{P}^2&3&2\end{array}\right]\; .
\end{equation} 
The last equivalence follows by repeated application of the equivalence $[\mathbb{P}^n\,|\,1]\cong \mathbb{P}^{n-1}$. In order to find the points $Y_i$ explicitly, we make a particularly simple choice for the polynomials $Q_a$, namely
\begin{equation}
 Q_1=y_1^3-y_0^3\;,\quad Q_2=y_2^2-y_0^2\;,\quad Q_a=y_a\quad\;\mbox{for}\quad\; a=3,\ldots ,6\; .
\end{equation} 
Then, the six points are given by
\begin{equation}
 \{Y_i\}=\{[1:\alpha^q:(-1)^s:0:\cdots :0]\in\mathbb{P}^6\,~|~\, q=0,1,2\,,\; s=0,1\}\; ,
\end{equation}
where $\alpha=\exp(2\pi i/3)$.  For the projective and affine coordinate ring of these points we have
\begin{align}
 S&=\frac{\mathbb{C}[y_0,\ldots ,y_6]}{\langle y_1^3-y_0^3,y_2^2-y_0^2,y_3,y_4,y_5,y_6\rangle}\hspace{12mm}\cong\frac{\mathbb{C}[y_0,y_1,y_2]}{\langle y_1^3-y_0^3,y_2^2-y_0^2\rangle} \\
 A&=\frac{\mathbb{C}[y_0,\ldots ,y_6]}{\langle y_1^3-y_0^3,y_2^2-y_0^2,y_3,y_4,y_5,y_6,y_0-1\rangle}\cong\frac{\mathbb{C}[y_1,y_2]}{\langle y_1^3-1,y_2^2-1\rangle} \nonumber\\
 &={\rm Span}([1],[y_1],[y_1^2],[y_2],[y_1y_2],[y_1^2y_2])\; .
\end{align}
In the last expression the square brackets indicate the class in $A$ and we see explicitly that $A$ is six-dimensional. The existence of a basis of $A$ with monomial representatives is a general feature of such affine coordinate rings for zero-dimensional varieties, as we discuss in Section~\ref{sec:hilbert}.
\vskip 2mm
\noindent{\bf Example 2:} {\it A type II example with two projective factors}\\
The CICY manifold $X$ (with number $7888$ in the standard list~\cite{Candelas:1987kf,Green:1986ck}) is defined by the configuration matrix
\begin{equation}
 X\in\left[\begin{array}{l|ll}\mathbb{P}^1&0&2\\\mathbb{P}^4&4&1\end{array}\right]^{2,86}_{-168}\qquad
 \begin{array}{l}x_0,x_1\\y_0,\ldots ,y_4\end{array}\; . \label{confex2}
\end{equation} 
The single-wrapping Gromov-Witten invariant for the class associated to the $\mathbb{P}^1$ factor is $n_\C=4$ and the loci $Y_i$ of the four curves in $\mathbb{P}^4$ are described by the configuration matrix
\begin{equation}
 \{Y_1,Y_2,Y_3,Y_4\}\in\left[\begin{array}{l|llll}\mathbb{P}^4&4&1&1&1\end{array}\right]\cong \left[\begin{array}{l|l}\mathbb{P}^1&4\end{array}\right]\; .
\end{equation} 
For a simple choice of defining polynomials we can explicitly compute the four points.
\begin{equation}
 Q_1=y_1^4-y_0^4\;,\quad Q_a=y_a\;\mbox{for}\;a=2,3,4\qquad\Rightarrow\qquad Y_q=[1:i^{q-1}:0:0:0:0],~ q=0,1,2,3\; .
\end{equation}
The projective and affine coordinate rings of these four points are given by
\begin{align}
 S&=\frac{\mathbb{C}[y_0,\ldots ,y_4]}{\langle y_1^4-y_0^4,y_2,y_3,y_4\rangle}\hspace{12mm}\cong\frac{\mathbb{C}[y_0,y_1]}{\langle y_1^4-y_0^4\rangle}\\
 A&=\frac{\mathbb{C}[y_0,\ldots ,y_4]}{\langle y_1^4-y_0^4,y_2,y_3,y_4,y_0-1\rangle}\cong\frac{\mathbb{C}[y_1]}{\langle y_1^4-1\rangle}
      ={\rm Span}([1],[y_1],[y_1^2],[y_1^3])\; .
\end{align} 
\vskip 2mm

\noindent{\bf Example 3:} {\it A type I example with three projective factors}\\
For a more complicated type I example with three projective factors we consider the CICY $X$ (number $7804$ in the standard list~\cite{Candelas:1987kf,Green:1986ck})) with configuration matrix
\begin{equation}
 X\in\left[\begin{array}{c|ccc}\mathbb{P}^1&0&1&1\\\mathbb{P}^2&1&1&1\\\mathbb{P}^3&3&1&0\end{array}\right]^{3,57}_{-108}\qquad \begin{array}{l} x_0,x_1\\\tilde{y}_0,\tilde{y}_1,\tilde{y}_2\\y_0,y_1,y_2,y_3\end{array}\; . \label{confex3}
\end{equation} 
The single-wrapping Gromov-Witten invariant in the $\mathbb{P}^1$ direction is $n_\C=3$ and the loci $Y_i$ of these three curves in $\mathbb{P}^2\times\mathbb{P}^3$ are described by the configuration matrix
\begin{equation}
 \{Y_1,Y_2,Y_3\}\in\left[\begin{array}{l|lllll}\mathbb{P}^2&1&1&1&1&1\\\mathbb{P}^3&3&1&1&0&0\end{array}\right]\cong\left[\begin{array}{l|lll}\mathbb{P}^3&3&1&1\end{array}\right]\cong\left[\begin{array}{l|l}\mathbb{P}^1&3\end{array}\right]\; .
\end{equation} 
With simple defining equations
\begin{equation}
 Q_1=\tilde{y}_0y_1^3-\tilde{y}_0y_0^3\;,\quad Q_2=\tilde{y}_0y_2\;\quad Q_3=\tilde{y}_0y_3\;,\quad Q_4=\tilde{y}_1\;,\quad Q_5=\tilde{y}_2\; ,
\end{equation} 
the three points are explicitly given by
\begin{equation}
 Y_q=([1:0:0],~[1:\alpha^{q-1}:0:0]),~q=0,1,2\; ,
\end{equation}
where $\alpha=\exp(2\pi i/3)$. For the projective and affine coordinate rings of these points we have
\begin{align}
 S&=\frac{\mathbb{C}[\tilde{y}_0,\tilde{y}_1,\tilde{y}_2,y_0,y_1,y_2 ,y_3]}{\langle \tilde{y}_0y_1^3-\tilde{y}_0y_0^3,\tilde{y}_0y_2,\tilde{y}_0y_3,\tilde{y}_1,\tilde{y}_2\rangle}\hspace{24mm}
 \cong \frac{\mathbb{C}[y_0,y_1]}{\langle y_1^3-y_0^3\rangle}\\
 A&=\frac{\mathbb{C}[\tilde{y}_0,\tilde{y}_1,\tilde{y}_2,y_0,y_1,y_2 ,y_3]}{\langle \tilde{y}_0y_1^3-\tilde{y}_0y_0^3,\tilde{y}_0y_2,\tilde{y}_0y_3,\tilde{y}_1,\tilde{y}_2,\tilde{y}_0-1,y_0-1\rangle}
 \cong \frac{\mathbb{C}[y_1]}{\langle y_1^3-1\rangle}
 ={\rm Span}([1],[y_1],[y_1^2])\; .
\end{align} 
\vskip 2mm

\noindent{\bf Example 4:} {\it A type II example with three projective factors}\\
Our final example is a CICY $X$ (number $7881$ in the standard list~\cite{Candelas:1987kf,Green:1986ck}) with configuration matrix
\begin{equation}
 X\in\left[\begin{array}{c|cc} \mathbb{P}^1&0&2\\\mathbb{P}^1&2&0\\\mathbb{P}^3&3&1\end{array}\right]^{3,75}_{-144}\quad
 \begin{array}{l}x_0,x_1\\y_0,y_1\\\tilde{y}_0,\tilde{y}_1,\tilde{y}_2,\tilde{y}_3\end{array}\label{confex4}
\end{equation} 
and a single-wrapping genus zero Gromov-Witten invariant for the class associated to the first $\mathbb{P}^1$ factor of $n_\C=2$. The loci $Y_i\in\mathbb{P}^1\times\mathbb{P}^3$ of the two curves in the transverse space are described by the configuration matrix
\begin{equation}
 \{Y_1,Y_2\}\in\left[\begin{array}{c|cccc}\mathbb{P}^1&2&0&0&0\\\mathbb{P}^3&3&1&1&1\end{array}\right]\cong\left[\begin{array}{l|l}\mathbb{P}^1&2\end{array}\right]\; .
\end{equation} 
With simple choices for the defining equations, these two points are easily computed:
\begin{equation}
 Q_1=\tilde{y}_0^3y_1^2-\tilde{y}_0^3y_0^2\;,\quad Q_2=\tilde{y}_1\;,\quad Q_3=\tilde{y}_2\;,\quad Q_4=\tilde{y}_3\qquad\Rightarrow\qquad
 Y_\pm=([1:\pm 1],[1:0:0:0])\;.
\end{equation} 
The projective and affine coordinate rings of these two points are
\begin{align}
 S&=\frac{\mathbb{C}[y_0,y_1,\tilde{y}_0,\tilde{y}_1,\tilde{y}_2\tilde{y}_3]}{\langle \tilde{y}_0^3y_1^2-\tilde{y}_0^3y_0^2,\tilde{y}_1,\tilde{y}_2,\tilde{y}_3\rangle}\hspace{24mm}\cong
 \frac{\mathbb{C}[y_0,y_1]}{\langle y_1^2-y_0^2\rangle}\\
 A&=\frac{\mathbb{C}[y_0,y_1,\tilde{y}_0,\tilde{y}_1,\tilde{y}_2\tilde{y}_3]}{\langle \tilde{y}_0^3y_1^2-\tilde{y}_0^3y_0^2,\tilde{y}_1,\tilde{y}_2,\tilde{y}_3,y_0-1,\tilde{y}_0-1\rangle}\cong  \frac{\mathbb{C}[y_1]}{\langle y_1^2-1\rangle}={\rm Span}([1],[y_1])
 \end{align}

\section{The bundle}\label{sec:bundle}
Our next step is the construction of vector bundles $V\rightarrow X$ over the CICY manifolds introduced in the previous section. There are, of course, many ways to construct such bundles. Here we focus on three standard methods, namely, line bundle sums, extension bundles and monad bundles. We consider each of these classes in turn and discuss how they relate to the geometric method for instanton calculations outlined in Section~\ref{sec:general}.

\subsection{Line bundle sums}
Recall that we are working with CICY manifolds $X\subset{\cal A}$ in an ambient space of the form ${\cal A}=\mathbb{P}^1\times {\cal B}$, with ${\cal B}$ an arbitrary product of projective factors. Line bundles on $X$ are denoted by ${\cal O}_X(k,\hat{k})$, where $k$ is the degree in the $\mathbb{P}^1$ direction and $\hat{k}$ the multi-degree in the factors of ${\cal B}$. As our vector bundle we take a rank $r\leq 8$ line bundle sum
\begin{equation}
 V=\bigoplus_{a=1}^r{\cal O}_X(k_a,\hat{k}_a)\; .
\end{equation} 
As usual, we impose $c_1(V)=0$ so that an embedding into $E_8$ is possible and this is equivalent to
\begin{equation}
 c_1(V)=0\qquad\Longleftrightarrow\qquad \sum_{a=1}^rk_a=\sum_{a=1}^r\hat{k}_a=0\; . \label{c1cond}
\end{equation} 
To guarantee bundle supersymmetry we require that there is a locus in the Kahler moduli space where the slopes of all line bundles vanish. Finally, we require that the curve dual to $c_2(TX)-c_2(V)$ is in the Mori cone of $X$ so that there exist a supersymmetric, anomaly-free completion of the model. These conditions impose further constraints on the line bundle integers $k_a$ and $\hat{k}_a$ which can be easily worked out. We refrain from doing so as the details are not relevant for our discussion of instanton effects.\\[2mm]
From Eq.~\eqref{H0meth}, we need to consider the bundles $V_i=V|_{C_i}\otimes{\cal O}_{\mathbb{P}^1}(-1)$ in order to calculate the Pfaffians. Remembering that the curves $C_i$ are given by a point in ${\cal B}$ times the first $\mathbb{P}^1$ factor, these bundles are easily computed by restricting the line bundles to the degrees in the $\mathbb{P}^1$ direction.
\begin{equation} 
 V_i=V|_{C_i}\otimes{\cal O}_{\mathbb{P}^1}(-1)=\bigoplus_{a=1}^r{\cal O}_{\mathbb{P}^1}(k_a-1)\; .
\end{equation} 
Recall that the cohomology dimensions for line bundles on $\mathbb{P}^1$ is governed by the formulae
\begin{equation}\label{P1coh}
 h^0({\cal O}_{\mathbb{P}^1}(k))=\left\{\begin{array}{lll}k+1&\mbox{for}&k\geq 0\\0&\mbox{for}&k<0\end{array}\right.\; ,\qquad
 h^1({\cal O}_{\mathbb{P}^1}(k))=\left\{\begin{array}{lll}0&\mbox{for}&k\geq 0\\-k-1&\mbox{for}&k<0\end{array}\right.\; .
\end{equation} 
This implies immediately that
\begin{equation}
 h^0( V_i)=\sum_{\{a|k_a\geq0\}}k_a\; .
\end{equation} 
Combining this result with Eq.~\eqref{H0meth} and Eq.~\eqref{c1cond} leads to a very simple criterion for the vanishing of the instanton superpotential.
\begin{equation}
 \mbox{At least one }k_a\neq 0\qquad\Longleftrightarrow\qquad W_\C=0\; . \label{lbscrit}
\end{equation}
 In other words, the only cases which lead to non-vanishing instanton superpotentials are the ones where all line bundles restrict trivially to the curves $C_i$.\\[2mm]
In conclusion, for line bundle sums we have a rather simple and satisfactory criterion for the vanishing of the instanton superpotential $W_\C$. However, note that line bundle sums typically do have moduli and represent special ``split loci" in a moduli space of bundles which generically have a non-Abelian structure group. The vanishing of $W_\C$ for a line bundle sum does not necessarily imply that $W_\C$ remains zero once we move away from the line bundle locus in moduli space. To address this problem we need to consider other bundle constructions which allow for non-Abelian structure groups.

\subsection{Monad and extension bundles}
Extensions and monads are two standard methods to construct bundles with a non-Abelian structure group. We would now like to consider these two classes and summarize how they relate to instanton superpotential calculations.\\[2mm]
The monad and extension bundles will be built from two line bundle sums
\begin{equation}
A=\bigoplus_{\alpha=1}^{r_A}{\cal O}_X(a_\alpha,\hat{a}_\alpha)\; ,\qquad B=\bigoplus_{\beta=1}^{r_B}{\cal O}_X(b_\beta,\hat{b}_\beta)\; .
\end{equation}
where we recall that the first entries $a_\alpha$, $b_\beta$ denote the degree in the $\mathbb{P}^1$ direction and $\hat{a}_\alpha$, $\hat{b}_\beta$ are the multi-degrees in the transverse space ${\cal B}$. It is also useful to introduce the restrictions of these line bundle sums to the curves $C_i$, tensored with $\OP(-1)$, since these bundles determine the properties of the instantons.
\begin{equation} 
 A_i:=A|_{C_i}\otimes{\cal O}_{\mathbb{P}^1}(-1)=\bigoplus_\alpha {\cal O}_{\mathbb{P}^1}(a_\alpha-1)\;,\qquad
 B_i:=B|_{C_i}\otimes{\cal O}_{\mathbb{P}^1}(-1)=\bigoplus_\beta {\cal O}_{\mathbb{P}^1}(b_\beta-1)\; . \label{ABdef}
\end{equation}
In terms of the above line bundle sums, monad and extension bundles $V\rightarrow X$ are defined by short exact sequences and their properties are summarized in the following table.
\begin{center}
\begin{tabular}{|c|c|c|}\hline
&monads&extensions\\\hline\hline
sequence&$0\longrightarrow V\longrightarrow A\stackrel{F}{\longrightarrow} B\longrightarrow 0$&$0\longrightarrow A\longrightarrow V\longrightarrow B\longrightarrow 0$\\\hline
map&$F\in H^0(B\otimes A^*)$&$ \delta\in {\rm Ext}^1(B,A)\cong H^1(A\otimes B^*)$\\\hline
${\rm rk}(V)$&$r_A-r_B$&$r_A+r_B$\\\hline
$c_1(V)$&$c_1(A)-c_1(B)$&$c_1(A)+c_1(B)$\\\hline
$H^0(V_i)$&${\rm Ker}\left( H^0(A_i)\stackrel{\delta_i}{\longrightarrow} H^0(B_i)\right)$&${\rm Ker}\left( H^0(B_i)\stackrel{\delta_i}{\longrightarrow} H^1(A_i)\right)$\\\hline
\end{tabular}
\end{center}
For either construction, we should impose that ${\rm r}(V)\leq 8$ and $c_1(V)=0$ which leads to certain constraints on the line bundle integers. Further constraints arise from bundle superymmetry and the anomaly conditions but there is no need to discuss these in detail.
 It is worth noting that the cohomology dimensions which appear in the last row can be easily computed from Eq.~\eqref{P1coh} and are given by
\begin{equation}
 h^0(A_i)=\sum_{\{\alpha|a_\alpha\geq0\}}a_\alpha\;,\qquad h^0(B_i)=\sum_{\{\beta|b_\beta\geq 0\}}b_\beta\;,\qquad h^1(A_i)=-\sum_{\{\alpha|a_\alpha\leq 0\}}a_\alpha\; . 
 \label{hAB}
\end{equation} 
In analogy with Eq.~\eqref{ABdef}, we also introduce the restriction
\begin{equation}
 V_i:=V|_{C_i}\otimes \OP(-1)
\end{equation}
of $V$ to the curve $C_i$. Since the index $\chi(V_i)=c_1(V)=0$ vanishes from the index theorem we conclude that
\begin{equation}
 h^0(V_i)=h^1(V_i)\; . \label{Veq}
\end{equation} 
Next, consider the long exact sequence in cohomology associated to the monad sequence restricted to $C_i$.
\begin{equation}
 \begin{array}{lllllllll}
  0&\longrightarrow&H^0(V_i)&\longrightarrow&H^0(A_i)&\stackrel{\delta_i}{\longrightarrow}&H^0(B_i)&&\\
  &\longrightarrow&H^1(V_i)&\longrightarrow&H^1(A_i)&\longrightarrow&H^1(B_i)&\longrightarrow&0
  \end{array}\; .\label{P1seqmon}
\end{equation}
Combining this sequence with the equality~\eqref{Veq} shows that whenever $h^0(A_i)\neq h^0(B_i)$ we must have $h^0(V_i)\neq 0$. The analogous long exact sequence for extensions
\begin{equation}
 \begin{array}{lllllllll}
  0&\longrightarrow&H^0(A_i)&\longrightarrow&H^0(V_i)&\longrightarrow&H^0(B_i)&&\\
  &\stackrel{\delta_i}{\longrightarrow}&H^1(A_i)&\longrightarrow&H^1(V_i)&\longrightarrow&H^1(B_i)&\longrightarrow&0
  \end{array}\; ,\label{P1seq}
\end{equation}
together with Eq.~\eqref{Veq} leads to a similar conclusion. For $h^0(B_i)\neq h^1(A_i)$ we must necessarily have $h^0(V_i)\neq 0$. Combining these observations with the criterion~\eqref{H0meth} then proves the simple vanishing statement
\begin{equation}
 \left\{\begin{array}{ll}h^0(A_i)\neq h^0(B_i)&\mbox{for monads}\\h^0(B_i)\neq h^1(A_i)&\mbox{for extensions}\end{array}\right\}\qquad\Rightarrow\qquad W_\C=0\; . \label{WC0}
\end{equation} 
In other words, all cases with a non-zero instanton superpotential must necessarily satisfy
\begin{equation}
 h^0(A_i)=h^0(B_i)\;\mbox{for monads}\;,\qquad h^0(B_i)=h^1(A_i)\;\mbox{for extensions}\; , \label{WCother}
\end{equation} 
and, from now on, we will assume these relations are satisfied. Then, we can think of the maps $\delta_i$ as square matrices and introduce the determinants
\begin{equation}
 f_i={\rm det}(\delta_i)\; .
\end{equation} 
Clearly, $H^0(V_i)\neq 0$ if and only if $f_i=0$ and, hence, the $f_i$ are the functions of the same name which we have introduced in Section~\ref{sec:general} and which enter the criterion~\eqref{fcrit}. The maps $\delta_i$ can be computed by restricting the monad map $F$ or the extension map $\delta$ to the cycle $C_i$, and then working out the induced map on cohomology. In cases where the monad and extension maps descend from ambient space polynomials, on which we focus here, this always leads to functions $f_i$ which can be expressed as
\begin{equation}
 f_i=f|_{\C_i}\;,\qquad f\in \mathbb{C}[y]_k\; , \label{fdef}
 \end{equation}
that is, as a restriction to $C_i$ of polynomials $f$ with a certain multi-degree $k$ in the directions of the transverse space ${\cal B}$. Different choices of $f$ with this multi-degree reflect different points in the bundle moduli space - we can think of the coefficients of a general $f\in\mathbb{C}[y]_k$ as (some of the) bundle moduli $b$. Note that this considerably simplifies the structure of the discussion. All we need to know is the multi-degree $k$ in order to determine the crucial maps $f_i$. It can be computed from the line bundle integers $\hat{a}_\alpha$ and $\hat{b}_\beta$ but the precise relation depends on the case. Our subsequent discussion is largely independent  of these details and merely starts with Eq.~\eqref{fdef}. Some examples of the relation between $k$ and the line bundle integers are provided in Section~\ref{sec:ex2}.

\section{Coordinate rings and Hilbert functions}\label{sec:hilbert}
In this section we review some basic mathematical facts about zero-dimensional varieties and their coordinate rings and Hilbert functions. A useful mathematical reference for some of this material is \cite{cox}.\\[2mm]
We briefly recall the algebraic set-up which we have already introduced in Section~\ref{sec:setup}. For a product ${\cal B}=\mathbb{P}^{n_1}\times\cdots\times\mathbb{P}^{n_m}$ of projective spaces with homogeneous coordinates $y=(y_{\alpha,0},\ldots ,y_{\alpha,n_\alpha})_{\alpha=1,\ldots ,m}$ we have the associated multi-graded coordinate ring $\mathbb{C}[y]$, with multi-degrees denoted by $k=(k_1,\ldots ,k_m)$. It is also useful to introduce the standard open patch $U_0$ of ${\cal B}$ where all $y_{\alpha,0}\neq 0$. Assume we have a  zero-dimensional variety consisting of a finite number of points $\{Y_1,\ldots ,Y_n\}\subset {\cal B}$. In the context of instanton calculations, these points are of course the loci of the isolated, genus-zero curves in the transverse space ${\cal B}$. We can think of this point set as a projective sub-variety of ${\cal B}$  which is then described by a projective ideal $I\subset \mathbb{C}[y]$. Alternatively, if all points $Y_i$ are contained in $U_0$ we can think of it as an affine variety whose associated ideal $J=I+\langle y_{1,0}-1,\ldots, y_{m,0}-1\rangle$ is obtained from $I$ by adding the localising polynomials $y_{\alpha,0}-1$.\\[2mm]
The map $r:\mathbb{C}[y]\rightarrow S$ introduced in Eq.~\eqref{SAmap} is defined by $r(f)=[f]$, that is, it takes the class of a polynomial within $S=\mathbb{C}[y]/I$. The map $\ell :S\rightarrow A$ is the localisation map which, in practice, amounts to setting all $y_{\alpha,0}=1$. The affine coordinate ring $A=\mathbb{C}[y]/J$ is finite-dimensional and its dimension ${\rm dim}(A)=n$ equals the number of points it describes. It is also known~\cite{cox} that is has a basis with monomial representatives.

\subsection{Hilbert functions} \label{sec:hf}
The rings $\mathbb{C}[y]$ and $S$ are multi-graded and they have standard Hilbert functions. For the ring $S$, the Hilbert function $h_S$ and the Hilbert series $H_S$ are defined by
\begin{equation}
 h_S(k)={\rm dim} (S_k)\;,\qquad H_S(t)=\sum_k h_S(k) t^k\; .
\end{equation} 
In other words, the Hilbert function gives the dimension of each multi-degree part $S_k$ of $S$ while the Hilbert series is simply the generating series for the Hilbert function (where $t^k=t_1^{k_1}\cdots t_m^{k_m}$). For sufficiently large degrees $k$, the Hilbert function is described by a polynomial -- the so-called Hilbert polynomial -- whose degree equals the dimension of the associated variety. Since we are concerned with a zero-dimensional variety, the Hilbert function becomes a constant for large $k$ which is, in fact, equal to the number $n$ of points.
\begin{equation}
 h_S(k)\rightarrow n\quad\mbox{for}\quad k\gg 1\; .
\end{equation} 
There are standard methods to compute the Hilbert function $h_S$, in particular by using syzygies in cases where the variety is a complete intersection. Since $h_S$ is not the main object of interest for our discussion we refrain from providing further details (see, for example, Ref.~\cite{harris}).\\[2mm]
The affine ring $A$ is not graded but filtered, with the filtration induced by the sub-algebras $A_{\leq k}$ of elements with multi-degree less or equal than $k$. The affine Hilbert function and series for $A$ are somewhat less common and are defined by
\begin{equation}
 h_A(k):={\rm dim}(A_{\leq k})\;,\qquad H_A(t)=\sum_kh_A(k)t^k\; .
\end{equation} 
From Eq.~\eqref{SAmap}, we have $A_{\leq k}=\ell(S_k)$, which implies that
\begin{equation}
 h_S(k)\geq h_A(k)
\end{equation}
for all $k$.  Unfortunately, equality does not always hold since the map $\ell|_{S_k}$ is not necessarily injective. Since $A$ has a finite basis with monomial representatives, it is clear that $h_A$ has the same asymptotic behaviour as $h_S$, namely
\begin{equation}
 h_A(k)\rightarrow n\quad\mbox{for}\quad k\gg1\; . \label{hAlimit}
\end{equation} 
How can the affine Hilbert function $h_A$ be computed? The following provides a basic algorithm.
\begin{enumerate}
\item Compute a Groebner basis $G=(g_i)$ of $J$.
\item Compute a monomial basis $B=(b_i)$ of (class representatives of) $A$ by collecting all monomials not contained in $\langle {\rm LT}(g_i)\rangle$, where ${\rm LT}(g_i)$ denotes the leading term of $g_i$ as induced by the ordering chosen in the Groebner basis computation. 
\item Select a monomial basis $(m_i)$ of $\mathbb{C}[y]_k$ and compute its remainders $m_i^G$ relative to the Groebner basis $G$. These remainders are linear combinations of the basis $B$.
\item Find the dimension of the space spanned by the remainders $m_i^G$. This dimension equals $h_A(k)$. 
\end{enumerate}

\subsection{Examples}
\begin{figure}[t]
\begin{center}
\includegraphics[width=6cm]{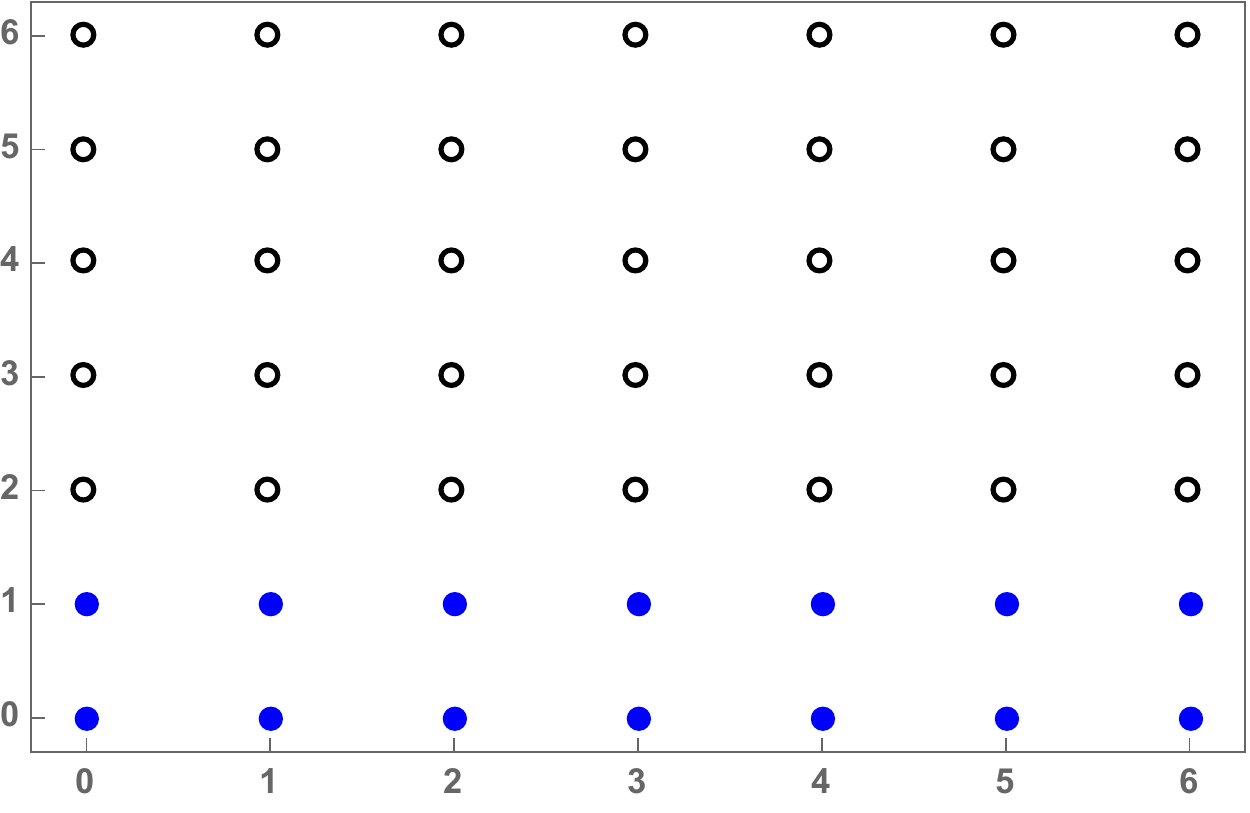}\hskip 1cm
\includegraphics[width=6cm]{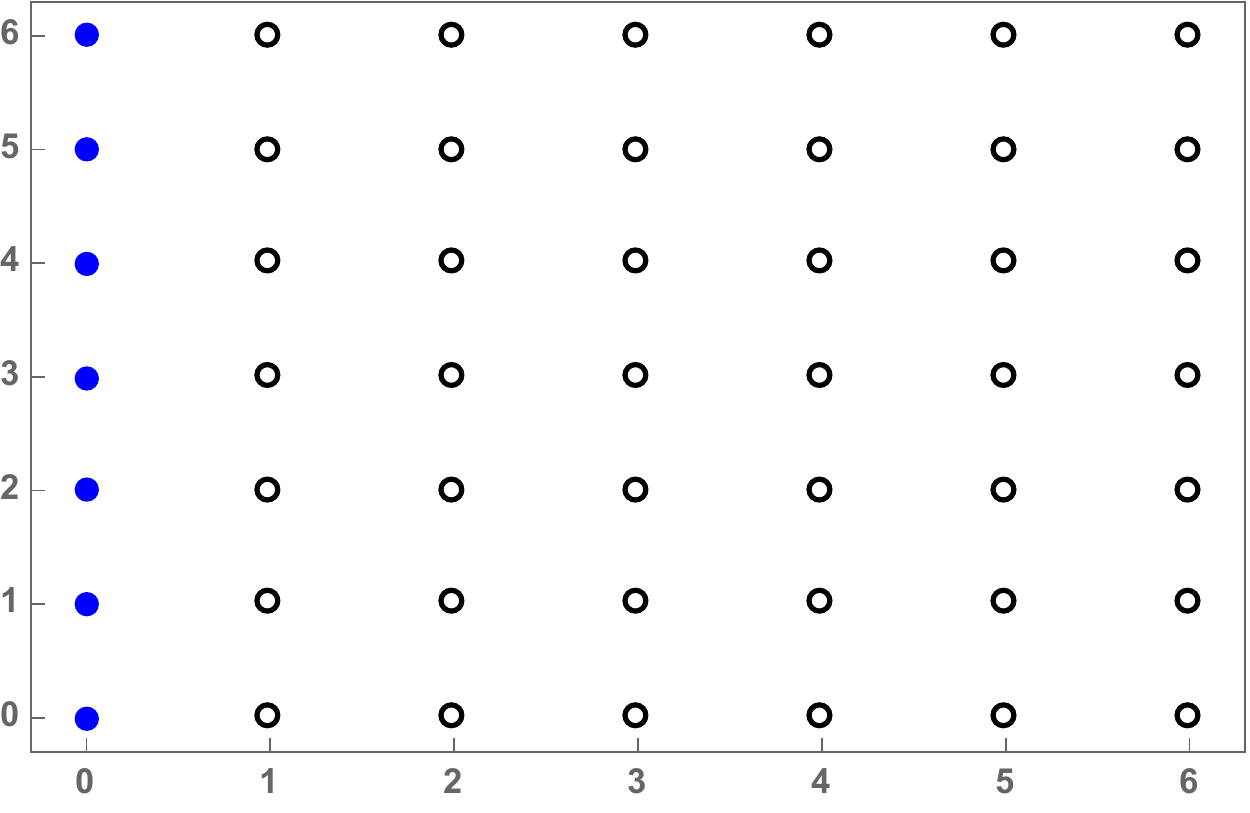}
\caption{\small\sf Results for affine Hilbert function $h_A(k)$ in the $k=(k_1,k_2)$ plane (with $k_1$ on the horizontal axis and $k_2$ on the vertical axis) for Example 3 with $n_\C=3$ (left) and Example 4 with $n_\C=2$ (right). Blue points indicate degrees for which $h_A(k)<n_\C$ and empty points satisfy $h_A(k)=n_\C$.}
\label{fig:ex3}
\end{center}
\end{figure}

Let us illustrate Hilbert functions and their computation by continuing with the example from Section~\ref{sec:examples}.\\[2mm]
\noindent {\bf Example 1:} Recall that this examples involves six points in $\mathbb{P}^6$ described by the configuration matrix
\begin{equation}
 \{Y_1,\ldots ,Y_6\}\in \left[\begin{array}{l|llllll}\mathbb{P}^6&3&2&1&1&1&1\end{array}\right]\cong\left[\begin{array}{l|ll}\mathbb{P}^2&3&2\end{array}\right]\; ,
\end{equation} 
and with coordinate rings
\begin{equation}
 S\cong\frac{\mathbb{C}[y_0,y_1,y_2]}{\langle y_1^3-y_0^3,y_2^2-y_0^2\rangle}\;,\qquad
 A\cong\frac{\mathbb{C}[y_1,y_2]}{\langle y_1^3-1,y_2^2-1\rangle} ={\rm Span}([1],[y_1],[y_1^2],[y_2],[y_1y_2],[y_1^2y_2])\; . \label{ex1SA}
\end{equation}
Using standard methods, the Hilbert series and Hilbert function for $S$ are obtained as
\begin{equation}
 H_S(t_1)=\frac{1+2t_1+2t_1^2+t_1^3}{1-t_1}\qquad\Rightarrow\qquad h_S(k)=\left\{\begin{array}{lll}2k+1&\mbox{for}&k<3\\6&\mbox{for}&k\geq 3\end{array}\right.\; .\label{hfex1}
\end{equation} 
To compute the affine Hilbert function we can follow the above algorithm. First we need to compute a Groebner basis $G$ for the ideal
\begin{align}
J=\langle y_1^3-y_0^3,y_2^2-y_0^2,y_0-1\rangle\,.
\end{align}
In lexicographic ordering, the Groebner basis and its leading terms are
\begin{align}
\label{eq:GroebnerBasisExample1}
G=(y_0-1,y_1^3-1,y_2^2-1)\quad\Rightarrow\quad\langle\text{LT}(g_i)\rangle=\langle y_0,y_1^3,y_2^2 \rangle\,.
\end{align}
Collecting terms not contained in $\langle\text{LT}(g_i)\rangle$, we find
\begin{align}
B=(1,y_1,y_1^2,y_2,y_1y_2,y_1^2y_2)\,,
\end{align}
and this is indeed the monomial basis for $A$ given in Eq.~\eqref{ex1SA}:  Next, we compute the monomial basis and its remainders. Let us look at $k=2$. A monomial basis for $\mathbb{C}[y]_2$ is simply
\begin{align}
(m_i)=(y_0^2,y_0y_1,y_0y_2,y_1^2,y_1y_2,y_2^2)\,.
\end{align}
Reducing this modulo \eqref{eq:GroebnerBasisExample1}, we find
\begin{align}
(m_i^G)=(1,y_1,y_2,y_1^2,y_1y_2,1)\,.
\end{align}
Since the space spanned by the remainders is five-dimensional we have $h_A(2)=5$. Continuing along those lines it is straightforward to verify that $h_A=h_S$, so in this case the two Hilbert functions coincide.\\[2mm]
\noindent {\bf Example 2:} This example involves four points in $\mathbb{P}^4$ with configuration matrix
\begin{equation}
 \{Y_1,Y_2,Y_3,Y_4\}\in\left[\begin{array}{l|llll}\mathbb{P}^4&4&1&1&1\end{array}\right]\cong \left[\begin{array}{l|l}\mathbb{P}^1&4\end{array}\right]\; .
\end{equation} 
and coordinate rings
\begin{equation}
 S\cong\frac{\mathbb{C}[y_0,y_1]}{\langle y_1^4-y_0^4\rangle}\;,\qquad
 A\cong\frac{\mathbb{C}[y_1]}{\langle y_1^4-1\rangle}={\rm Span}([1],[y_1],[y_1^2],[y_1^3])\; . \label{ex2SA}
\end{equation} 
The Hilbert series and function for $S$ are given by
\begin{equation}
 H_S(t_1)=\frac{1+t_1+t_1^2+t_1^3}{1-t_1}\qquad\Rightarrow\qquad h_S(k)=\left\{\begin{array}{lll}k+1&\mbox{for}&k<3\\4&\mbox{for}&k\geq 3\end{array}\right.\label{hfex2}
 \end{equation}
A quick inspection of the monomial basis for $A$ in Eq.~\eqref{ex2SA} shows that $h_A=h_S$, so again the Hilbert functions coincide.\\[2mm]
\noindent {\bf Example 3:} This example involves three points in ${\cal B}=\mathbb{P}^2\times\mathbb{P}^3$ described by a configuration matrix
\begin{equation}
 \{Y_1,Y_2,Y_3\}\in\left[\begin{array}{l|lllll}\mathbb{P}^2&1&1&1&1&1\\\mathbb{P}^3&3&1&1&0&0\end{array}\right]\; ,
\end{equation} 
and with associated coordinate rings
\begin{equation}
 S=\frac{\mathbb{C}[\tilde{y}_0,\tilde{y}_1,\tilde{y}_2,y_0,y_1,y_2 ,y_3]}{\langle \tilde{y}_0y_1^3-\tilde{y}_0y_0^3,\tilde{y}_0y_2,\tilde{y}_0y_3,\tilde{y}_1,\tilde{y}_2\rangle}\;,\qquad
 A=\frac{\mathbb{C}[\tilde{y}_0,\tilde{y}_1,\tilde{y}_2,y_0,y_1,y_2 ,y_3]}{\langle \tilde{y}_0y_1^3-\tilde{y}_0y_0^3,\tilde{y}_0y_2,\tilde{y}_0y_3,\tilde{y}_1,\tilde{y}_2,\tilde{y}_0-1,y_0-1\rangle}
\end{equation} 
The Hilbert series for $S$ as a bit more complicated
\begin{equation}
H_S(t)=\frac{1-t_1 t_2^5+2 t_1 t_2^4-t_1 t_2^3+t_1 t_2^2-2 t_1 t_2}{\left(1-t_1\right) \left(1-t_2\right){}^4}=1+t_1+4t_2+2t_1t_2+\cdots \label{ex3HF}
\end{equation}
and we have expanded only up to terms of degree $k\leq (1,1)$. The affine Hilbert function can be computed algorithmically, as discussed, and the result is schematically shown in Fig.~\ref{fig:ex3}. We note from Eq.~\eqref{ex3HF} that $h_S(0,1)=4$ while Fig.~\ref{fig:ex3} indicates that $h_A(0,1)<3$, in fact, $h_A(0,1)=2$. This is an example where the two Hilbert functions do not coincide - the map $\ell|_{S_{(0,1)}}$ is not injective.\\[2mm]
\noindent {\bf Example 4:} For this example, we have two points in ${\cal B}=\mathbb{P}^1\times\mathbb{P}^3$ with configuration matrix
\begin{equation}
 \{Y_1,Y_2\}\in\left[\begin{array}{c|cccc}\mathbb{P}^1&2&0&0&0\\\mathbb{P}^3&3&1&1&1\end{array}\right]\; ,
\end{equation} 
and coordinate rings
\begin{equation}
 S=\frac{\mathbb{C}[y_0,y_1,\tilde{y}_0,\tilde{y}_1,\tilde{y}_2\tilde{y}_3]}{\langle \tilde{y}_0^3y_1^2-\tilde{y}_0^3y_0^2,\tilde{y}_1,\tilde{y}_2,\tilde{y}_3\rangle}\;,\qquad
 A=\frac{\mathbb{C}[y_0,y_1,\tilde{y}_0,\tilde{y}_1,\tilde{y}_2\tilde{y}_3]}{\langle \tilde{y}_0^3y_1^2-\tilde{y}_0^3y_0^2,\tilde{y}_1,\tilde{y}_2,\tilde{y}_3,y_0-1,\tilde{y}_0-1\rangle}\; .
 \end{equation}
The Hilbert series for $S$ is given by
\begin{equation}
H_S(k)=\frac{1-t_1^2 t_2^3}{\left(1-t_1\right){}^2 \left(1-t_2\right)}=1+2t_1+t_2+2t_1t_2+\cdots
\end{equation}
and the results for the affine Hilbert function is schematically shown in Fig.~\ref{fig:ex3}. It turns out that in this case $h_A=h_S$ so the two Hilbert functions coincide.

\subsection{Evaluation from coordinate rings}
Recall that our goal is to use the criterion~\eqref{fcrit} for the non-vanishing of the instanton superpotential. This requires us to work out the functions $f_i$ which are proportional to the Pfaffians. We have seen in Eq.~\eqref{fdef} that they can be obtained from $f_i=f(Y_i)$, that is, by evaluating functions $f\in\mathbb{C}[y]_k$ of a certain multi-degree $k$ at the loci $Y_i$ of the curves $C_i$. This is straightforward in principle but might not be easy to carry out in practice since the points $Y_i$ may be hard to compute. 
We will now 
%briefly review 
propose an alternative method to calculate $f_i$ which does not rely on the explicit knowledge of the points $Y_i$ but uses the affine coordinate ring $A$ instead. 
%For more details, see Ref.~\cite{cox}.\\[2mm]
See Ref.~\cite{cox} for mathematical details underlying this approach.\\[2mm]
First recall that the affine coordinate ring $A$, associated to the point set $\{Y_1,\ldots ,Y_n\}$, is a finite-dimensional vector space of dimension $n$. We can define a linear map by
\begin{equation}
 \mu:\mathbb{C}[y]\rightarrow {\rm End}(A)\;,\qquad \mu(f)(a):=[f]\,a\; ,
\end{equation} 
where $[f]=\ell\circ r(f)$ is the class of the polynomial $f$ in $A$. Hence, for every polynomial $f\in\mathbb{C}[y]$ the image $\mu(f)$ is a linear map on $A$ which acts simply by the multiplication in the ring $A$. Since the ring multiplication is commutative, we have
\begin{equation}
 \mu(f) \mu(\tilde{f})= \mu(\tilde{f}) \mu(f)
\end{equation}
for all $f,\tilde{f}\in\mathbb{C}[y]$. In other words, all linear maps on $A$ obtained in this way commute with each other. The main mathematical statement we will be relying on is the following~\cite{cox}:
\begin{equation}
 \{f(Y_1),\ldots ,f(Y_n)\}=\{\mbox{eigenvalues of }\mu(f)\}\; .
\end{equation} 
This means, the crucial quantities $f_i=f(Y_i)$ proportional to the Pfaffians are given by the eigenvalues of the linear map $\mu(f):A\rightarrow A$. Moreover, all maps $\mu(f)$ obtained for $f$ ranging in $\mathbb{C}[y]_k$ commute and, hence, can be simultaneously diagonalised. \\[2mm]
This discussion allows us to re-formulate our original problem of linear (in)dependence of $f_i$ in terms of the properties of polynomials in the coordinate ring $A$. 
These properties can be studied using standard methods of commutative algebra and Hilbert series. As a result the criterion for a non-vanishing superpotential
can be stated using the Hilbert function as will be considered in the next section. 

\section{A Hilbert function condition for instantons} \label{sec:crit}
We are now ready to combine our various observations and formulate a condition for a non-zero instanton superpotential $W_\C$, based on the affine Hilbert function.  After stating the condition in general, we apply it to a range of examples.

\subsection{The general condition}
From our main criterion~\eqref{fcrit}, we need to decide whether or not the quantities $f_i=f(Y_i)$, where $i=1,\ldots, n_\C$, viewed as functions of bundle moduli $b$, are linearly independent. A practical way to re-formulate this is to choose a basis $(f_I)_{I=1,\ldots ,N}$ of $\mathbb{C}[y]_k$ and consider the $N\times n_\C$ matrix $M_{Ii}=f_I(Y_i)$. In terms of this matrix, the criterion~\eqref{fcrit} can be re-formulated as
\begin{equation}
 {\rm rk}(M)=n_\C\quad\Rightarrow\quad W_\C\neq 0\;. \label{fcrit2}
\end{equation} 
Let us point out that here it is assumed that the polynomial $f \in \mathbb{C}[y]_k$ is generic in the sense that we span the entire space $\mathbb{C}[y]_k$
as we vary its coefficients. In other words, $f$ can be expanded in the basis of $(f_I)_{I=1,\ldots ,N}$ with all basis elements present in the expansion. 
Otherwise, if only $N' <N$ basis elements appear in the expansion of $f$, 
we have to restrict $\mathbb{C}[y]_k$ to the subspace spanned by these basis elements. The matrix $M$ must now
be constructed using the basis elements $(f_I)_{I=1,\ldots ,N'}$ and is 
of the size  $N' \times n_\C$. However, the condition~\eqref{fcrit2} remains the same.\\[2mm]
Now consider the linear maps $\mu(f_I):A\rightarrow A$, as defined in the previous section. All these maps are simultaneously diagonalisable and the eigenvalues of $\mu(f_I)$ are precisely the entries $(M_{I1},\ldots ,M_{In_\C})$ of the $I^{\rm th}$ row of $M$.  Hence, it follows that
\begin{equation}
 {\rm rk}(M)={\rm dim}(\mu(\mathbb{C}[y]_k))={\rm dim}(\ell\circ r(\mathbb{C}[y]_k)={\rm dim}(A_{\leq k})=h_A(k)\; .
\end{equation}  
This means the criterion~\eqref{fcrit2} can be re-written in terms of the affine Hilbert function and then reads
\begin{equation}
 h_A(k)=n_\C\quad\Rightarrow\quad W_\C\neq 0\; . \label{fcritmain}
\end{equation}
This is our main result. We can use the affine Hilbert function of the coordinate ring $A$, which describes the locations of the curves $C_i$ in the transverse space, to decide whether the instanton superpotential $W_\C$ is non-zero. To do this, we have to determine the relevant multi-degree $k$ for the bundle $V$ in question. For common constructions, such as extension and monad bundles, this degree can usually be read off  from the defining data of the bundle. Some explicit examples of this are provided below. The simple conclusion is that, whenever the affine Hilbert function $h_A(k)$ takes its maximal value $n_\C$ (equal to the number of curves $C_i$), the instanton superpotential must be non-zero.  For cases with $h_A(k)<n_\C$ we cannot draw a definite conclusion and $W_\C$ can be zero or non-zero, depending on the undetermined constants $\lambda_i$ in Eq.~\eqref{instsp1}.
Note that the criterion~\eqref{fcritmain} does not depend on the precise locations of the points $Y_i$, which might be difficult to compute from the polynomial equations~\eqref{Qeqs}. 
It depends only on the Hilbert function of the coordinate ring $A$, which can be computed using methods of commutative algebra. \\[2mm]
The above result leads to a general picture for the non-vanishing of the instanton superpotential. First of all, we see from Eqs.~\eqref{WC0} that ``most" patterns which arise in common bundle constructions, such as monads and extensions, lead to a vanishing superpotential. However, there are specific patterns, characterised by the conditions~\eqref{WCother}, where the superpontial can be non-zero. For such cases, the answer depends on a multi-degree $k$ which can be extracted from the relevant bundle construction. The superpotential is non-zero if the Hilbert function criterion~\eqref{fcritmain} is satisfied. As Eq.~\eqref{hAlimit} shows, this criterion will be satisfied for sufficiently large $k$. This means, within the sub-class of bundles characterised by Eq.~\eqref{WCother}, a non-vanishing instanton superpotential is the ``typical" situation. We would now like to illustrate this general picture with a number of examples.

\subsection{Examples} \label{sec:ex2}
To set the scene, we indicate how the crucial multi-degree $k$ can be extracted from a given bundle construction. Consider a monad or extension bundle constructed from the line bundle sums
\begin{equation}
 A=\left(\begin{array}{clll}\pm c&0&\cdots &0\\\hat{a}_1&\hat{a}_2&\cdots&\hat{a}_{r_A}\end{array}\right)\;,\qquad
 B=\left(\begin{array}{clll}c&0&\cdots &0\\\hat{b}_1&\hat{b}_2&\cdots&\hat{b}_{r_B}\end{array}\right)
\end{equation} 
where each column contains the multi-degree of a line bundle, with the first row the degree in the $\mathbb{P}^1$ direction and the other rows the multi-degree in the transverse space ${\cal B}$. The upper sign in the $(1,1)$-entry of $A$ is for monads, the lower sign for extensions and $c$ is a positive integer. Note that for either case the condition~\eqref{WCother} is satisfied, so we have indeed a pattern where the instanton contribution can be non-vanishing. A quick calculation shows that the multi-degree $k$ for this pattern is given by
\begin{equation}
 k=\pm c(\hat{b}_1-\hat{a}_1)\; , \label{krel}
\end{equation}
with the upper sign for monads and the lower sign for extensions. Similar relations can be derived for other patterns. In the following, we will not be specific about this  relation but rather present our examples in terms of the multi-degree $k$. In this way, the results are applicable to a wide range of bundles, using equations such as~\eqref{krel}. We begin by revisiting our ``running" examples, introduced in Section~\eqref{sec:examples}.\\[2mm]
{\bf Example 1:} Our first example is for CICY manifold $7867$ in the ambient space ${\cal A}=\mathbb{P}^1\times\mathbb{P}^6$ defined by the configuration matrix~\eqref{confex1}. Its Picard number is $h^{1,1}(X)=2$ so $k$ is, in fact, just a single degree in this case. We have $n_\C=6$ curves and the Hilbert function $h_A=h_S$ , computed in  Eq.~\eqref{hfex1}, together with the criterion~\eqref{fcritmain}, shows that
\begin{equation}
 k\geq 3\quad\Rightarrow\quad W_\C\neq 0\; .
\end{equation} 
This illustrates our earlier statements that, within patterns of bundle constructions satisfying Eqs.~\eqref{WCother}, non-zero instanton superpotentials are common.\\[2mm]
{\bf Example 2:} This is CICY manifold $7888$ in the ambient space ${\cal A}=\mathbb{P}^1\times\mathbb{P}^4$ with configuration matrix~\eqref{confex2}. The Picard number is $h^{1,1}(X)=2$, so again $k$ is a single degree, and there are $n_\C=4$ curves. From the associated affine Hilbert function~\eqref{hfex2} ($h_A=h_S$ in this case) we conclude that
\begin{equation}
 k\geq 3\quad\Rightarrow\quad W_\C\neq 0\; .
\end{equation} 
\vskip 2mm
\noindent {\bf Example 3:} The CICY manifold $7804$ is defined in the ambient space ${\cal A}=\mathbb{P}^1\times\mathbb{P}^2\times\mathbb{P}^3$ with configuration matrix~\eqref{confex3}. There are $n_\C=3$ curves and since $h^{1,1}(X)=3$ we know that $k=(k_1,k_2)$ is a bi-degree. The affine Hilbert function in this case has been plotted in Fig.~\ref{fig:ex3} and it indicates that
\begin{equation}
 k_2\geq2\quad\Rightarrow\quad W_\C\neq 0\; .
\end{equation} 
Again, we see that the instanton superpotential is non-vanishing for ``most" bi-degrees $k=(k_1,k_2)$.\\[2mm]
{\bf Example 4:} CICY manifold $7881$ is defined in the ambient space ${\cal A}=\mathbb{P}^1\times\mathbb{P}^1\times\mathbb{P}^3$ with configuration matrix~\eqref{confex4}. It has $n_\C=2$ curves and Picard number $h^{1,1}(X)=3$, so that $k=(k_1,k_2)$ is a bi-degree. The associated affine Hilbert function, plotted in Fig.~\ref{fig:ex3}, shows that
\begin{equation}
 k_1\geq1\quad\Rightarrow\quad W_\C\neq 0\; .
\end{equation} 
\vskip 2mm
\noindent The above examples have been chosen for their relative simplicity, particularly a small number $n_\C$ of curves. We have computed the affine Hilbert function for a number of more complicated examples with Picard number $h^{1,1}(X)=3$, using the algorithm described in Section~\ref{sec:hf}. The results are shown in Table~\ref{tab:res}. For all cases, $k=(k_1,k_2)$ is a bi-degree and blue points in the figures correspond to bi-degrees with $h_A(k)<n_\C$ while empty points indicate $h_A(k)=n_\C$. From our main criterion~\eqref{fcritmain} all bi-degrees $k$ with empty points in those plots leads to a non-vanishing instanton superpotential $W_\C$.

\begin{table}
\begin{center}
\begin{tabular}{|m{5cm}|m{1cm}|m{1cm}|m{6cm}|}\hline
configuration&\#&$n_\C$&$h_A$\\\hline\hline
$\left[\begin{array}{l|llll}\mathbb{P}^1&1&1&0&0\\\mathbb{P}^3&2&0&1&1\\\mathbb{P}^3&0&2&1&1\end{array}\right]^{3,35}_{-64}$&$6771$&$32$&
\vskip 2mm\includegraphics[width=6cm]{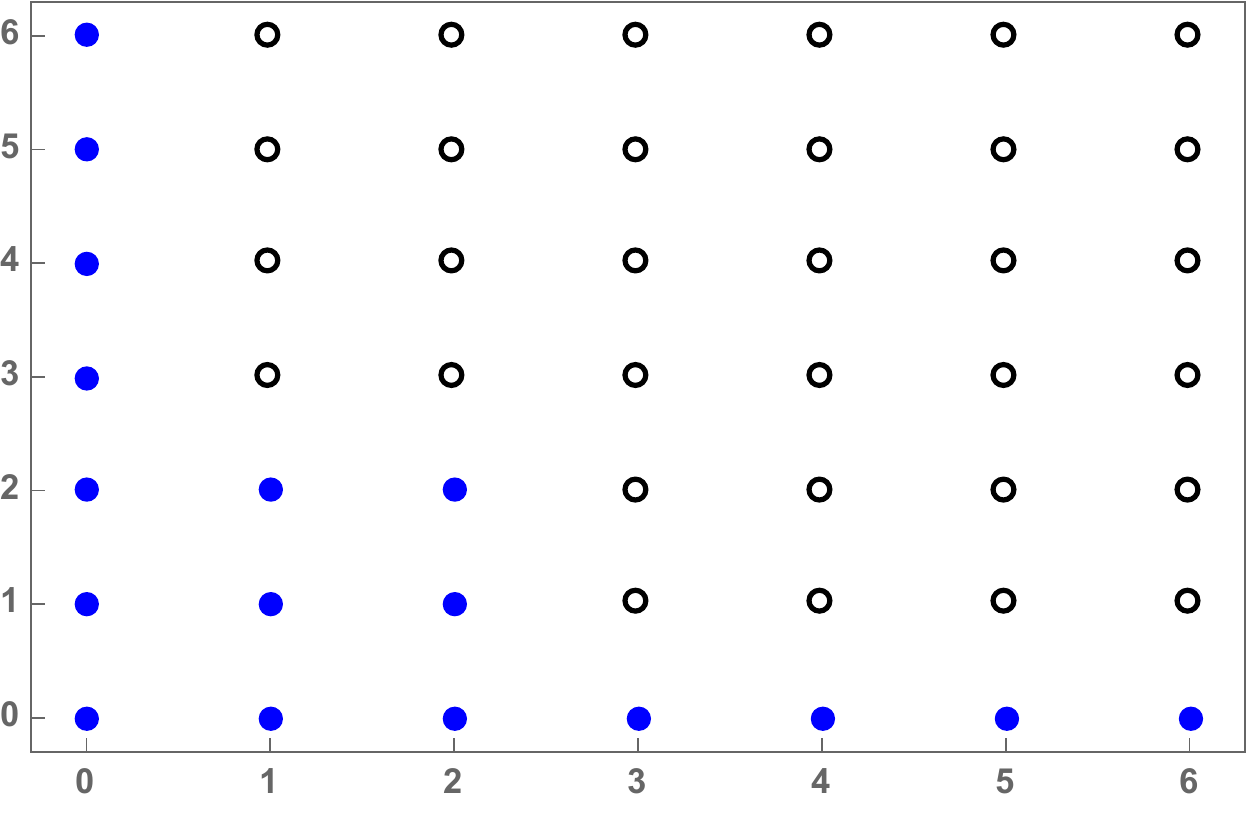}\\\hline
$\left[\begin{array}{l|llllll}\mathbb{P}^1&1&1&0&0&0&0\\\mathbb{P}^4&1&0&2&1&1&0\\\mathbb{P}^4&0&1&0&1&1&2\end{array}\right]^{3,39}_{-72}$&$7208$&$8$&
\vskip 2mm\includegraphics[width=6cm]{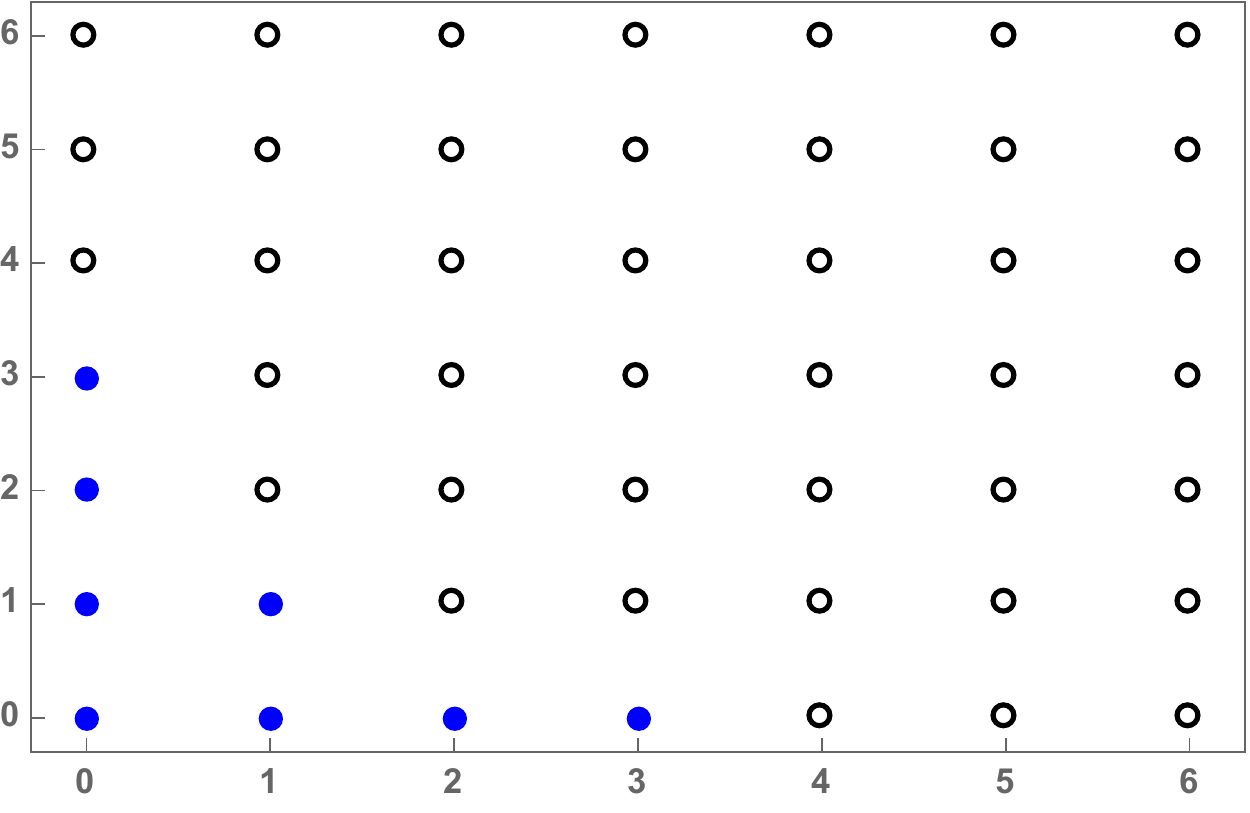}\\\hline
$\left[\begin{array}{l|llll}\mathbb{P}^1&2&0&0&0\\\mathbb{P}^2&0&1&1&1\\\mathbb{P}^4&2&1&1&1\end{array}\right]^{3,45}_{-84}$&$7585$&$24$&
\vskip 2mm\includegraphics[width=6cm]{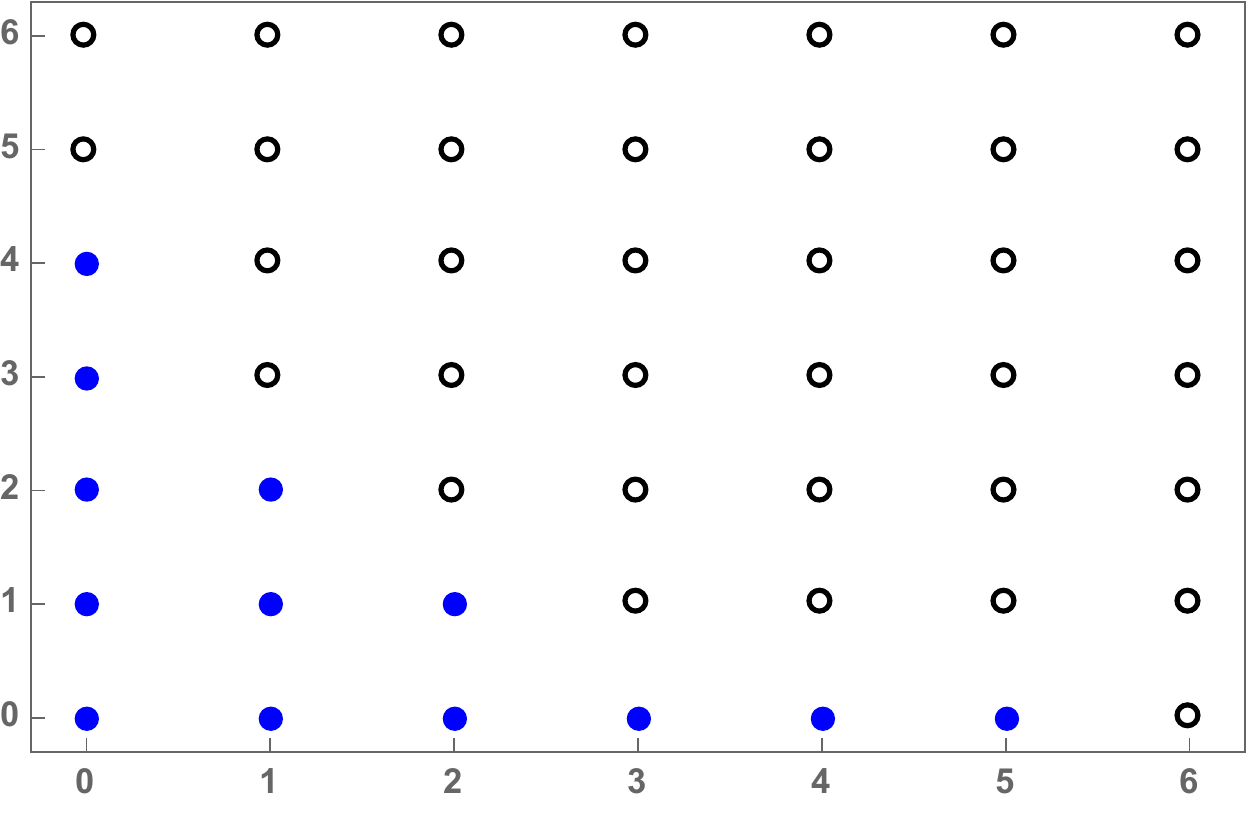}\\\hline
$\left[\begin{array}{l|lll}\mathbb{P}^1&1&1&0\\\mathbb{P}^2&2&0&1\\\mathbb{P}^3&1&2&1\end{array}\right]^{3,46}_{-86}$&$7610$&$32$&
\vskip 2mm\includegraphics[width=6cm]{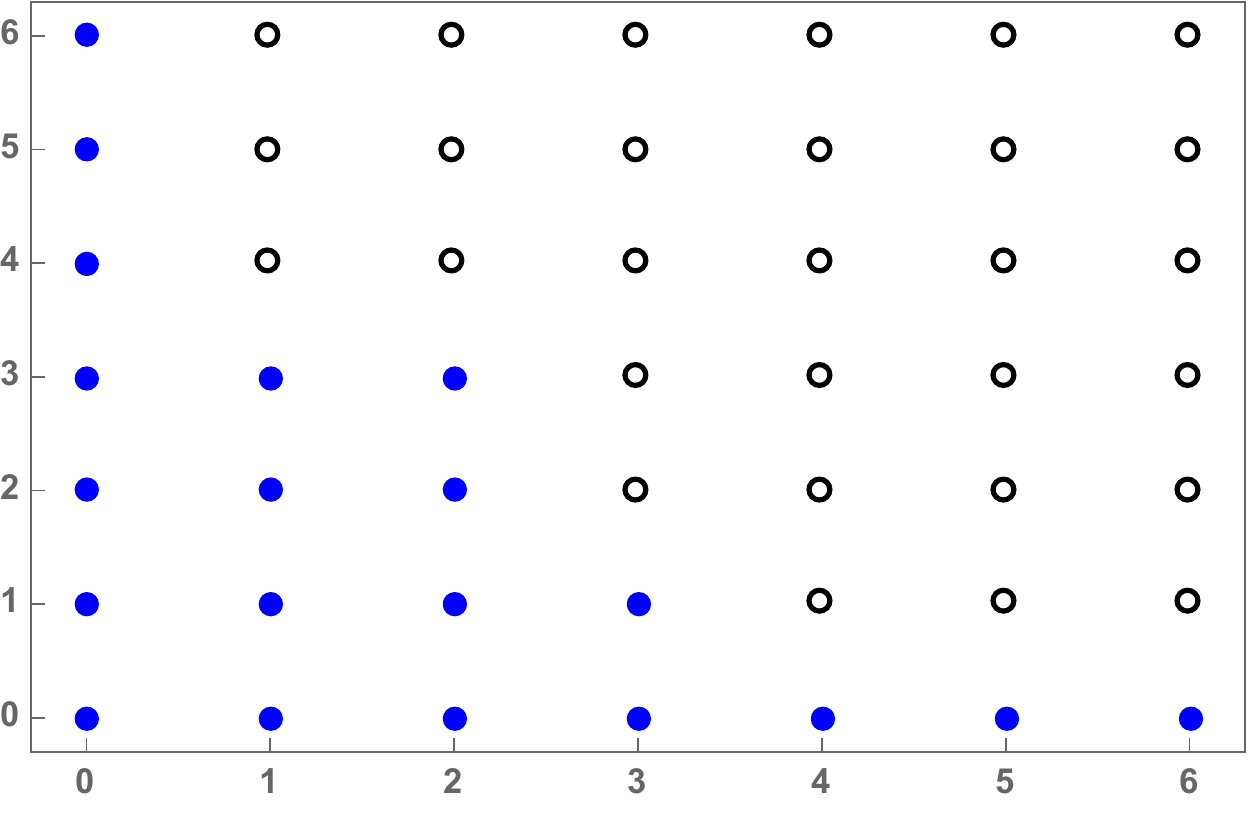}\\\hline
$\left[\begin{array}{l|llllll}\mathbb{P}^1&1&1&0&0&0&0\\\mathbb{P}^4&1&0&1&1&1&1\\\mathbb{P}^4&0&1&1&1&1&1\end{array}\right]^{3,47}_{-88}$&$7636$&$6$&
\vskip 2mm\includegraphics[width=6cm]{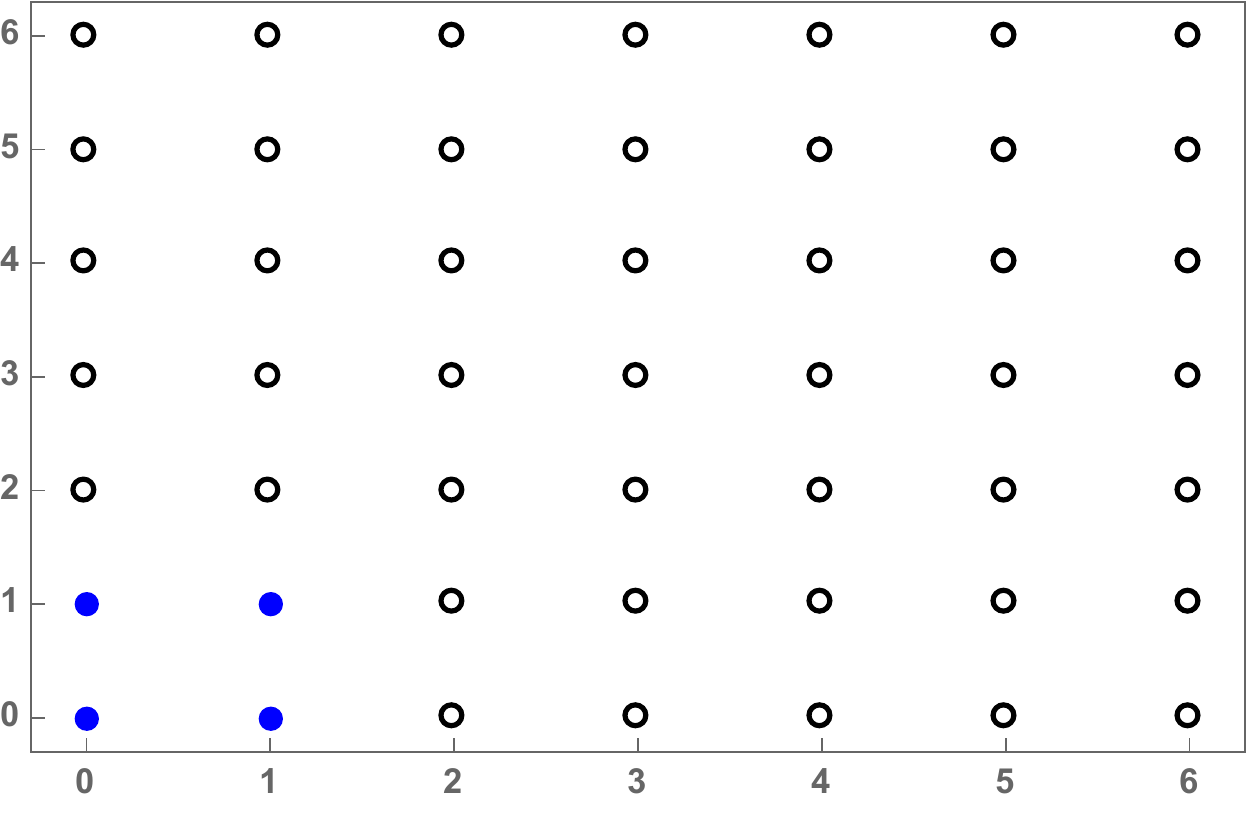}\\\hline
\end{tabular}
\end{center}
\vskip -3mm
\caption{\small\sf Affine Hilbert function $h_A(k)$ in the $k=(k_1,k_2)$ plane (with $k_1$ on the horizontal axis and $k_2$ on the vertical axis). Blue points have $h_A(k)<n_\C$ whereas empty points indicate that $h_A(k)=n_\C$.}\label{tab:res}
\end{table}

\section{Conclusion}\label{sec:conclusion}
In this paper, we have studied string instanton superpotentials for heterotic Calabi-Yau compactifications. Our main goal has been to find conditions for the vanishing/non-vanishing of the instanton superpotential $W_\C$ associated to a second homology class $\C$ of the Calabi-Yau manifold $X$. We have considered bundles $V\rightarrow X$ constructed from line bundle sums, monads and extensions.\\[2mm]
For line bundle sums we have found a simple criterion, Eq.~\eqref{lbscrit}, for the vanishing/non-vanishing of the instanton superpotential $W_\C$. It shows that non-vanishing instanton superpotentials for line bundle sums requires a special class of line bundles, which become trivial when restricted to the curves $C_i$, but that within this class, the superpotential is non-vanishing.\\[2mm]
For bundles with non-Abelian  structure groups, constructed from monads or extensions, the picture is somewhat more complicated. If certain cohomology dimensions of the constituent line bundles are not equal, as in Eq.~\eqref{WC0}, the instanton superpotential $W_\C$ vanishes. On the other hand, if these dimensions are equal, as in Eq.~\eqref{WCother}, the superpotential can be vanishing or non-vanishing.

In such cases, a criterion for non-vanishing superpotentials can be formulated in terms of an affine Hilbert function. This Hilbert function, $h_A$, is associated to the coordinate ring $A$ which describes the loci $Y_i$ of the $n_\C$ curves $C_i$ in a transverse space. What we have shown (see Eq.~\eqref{fcritmain}) is that whenever $h_A(k)=n_\C$, the instanton superpotential is non-zero. Here $k$ is a multi-degree which can be read off from the specific bundle construction. The asymptotic behaviour $h_A(k)\rightarrow n_\C$ for large $k$ means that a non-vanishing instanton superpotential is a common feature within this class.\\[2mm]
The first observation from these results is that non-vanishing instanton superpotentials are rare, in the sense that they require a specific pattern when constructing the bundle $V$. However, within the class of bundles following this pattern, the superpotential is either always non-zero (for line bundle sums) or it is frequently non-zero (for monads and extensions). These observations may well provide useful guidance for model-building, particularly in view of moduli stabilization.\\[2mm]
There are several interesting directions to pursue. The current formulation of our Hilbert function criterion depends on an ambient space of the form ${\cal A}=\mathbb{P}^1\times{\cal B}$, so that we can talk about the loci $Y_i$ of the curves $C_i$ in the transverse space ${\cal B}$ and introduce their associated coordinate ring $A$. It would be desirable to generalise this condition so that it can be applied to more general manifolds, possibly by introducing a coordinate ring associated to the union of all curves $C_i$. It is currently not clear how to formulate this.

Another deficit is that the criterion~\eqref{fcritmain} only works in one direction. If $h_A(k)<n_\C$ we are not able to draw a definite conclusion. Unfortunately, improving on this requires knowledge of the constants of proportionality $\lambda_i$ in the instanton superpotential~\eqref{instsp1}, which are hard to compute. Moreover, it is interesting that the condition for vanishing/non-vanishing instanton superpotentials depends on the degrees $k$ in the transverse space, while the compactness criterion of Bertolini-Plesser depends on the degree of the line bundles of the curve class under investigation. However, the two degrees are linked by anomaly cancellation and supersymmetry conditions, and we have shown for examples in which a GLSM description exists that the two condition give the same result. It would be interesting to show the equality of the two approaches algebraically.

Finally, our discussion has been limited to certain homology classes, related to $\mathbb{P}^1$ factors in the ambient space, and it would be desirable to remove this limitation.

\section*{Acknowledgements}
We would like to thank Lara Anderson, James Gray and Balazs Szendroi for helpful discussions.
The work of E.I.B.~is supported in part by the ARC Discovery Project DP200101944.
 A.L.~would like to acknowledge support by the STFC grant ST/L000474/1.  A.L.~and F.R.~would like to thank the University of Pennsylvania for hospitality. 
 A.L.~would also like to thank the Simons Center for Geometry and Physics, Stony Brook for hospitality. B.A.O is supported in part by DOE No. DE- SC0007901 and 
 SAS Account 020-0188-2-010202-6603-0338.

\end{document}